\documentclass[oneside,a4paper,12pt]{article}

\usepackage{graphicx,natbib,amssymb,lineno,hyperref}

\newcommand{\Mo}{$~\mathrm{M_\odot}$}
\def\astrobj#1{#1}

\newcommand\aj{{AJ }}%
\newcommand\apj{{ApJ }}%
\newcommand\aap{{A\&A }}%
\newcommand\aaps{{A\&AS }}%
\newcommand\mnras{{MNRAS }}%
\newcommand\pasp{{PASP }}%
\newcommand\bain{{Bull.~Astron.~Inst.~Netherlands }}%
\begin{document}

\begin{center}
\Large{\bf Period changes in six semi-detached Algol-type binaries}

 \medskip
\medskip

\large P. Zasche\small$^{a}$\footnote{Email: zasche@sirrah.troja.mff.cuni.cz}, {\large A. Liakos\small$^{b}$},
{\large M. Wolf\small$^{a}$}, {\large P. Niarchos\small$^{b}$}

\medskip
 \medskip

{\small$^{a}$ Astronomical Institute, Faculty of Mathematics and Physics,
 Charles University Prague, CZ-180 00 Praha 8, V Hole\v{s}ovi\v{c}k\'ach 2, Czech Republic}

\medskip

{\small$^{b}$ Department of Astrophysics, Astronomy and Mechanics, Faculty of Physics, National \& Kapodistrian
University of Athens, Athens, Greece}
\end{center}

\medskip
 \medskip
  \medskip

\noindent {\bf \underline{Abstract:}} Six semi-detached Algol-type binaries lacking a period analysis were
chosen to test for a presence of a third body. The $O-C$ diagrams of these binaries were analyzed with the
least-squares method by using all available times of minima. Also fourteen new minima, obtained from our
observations, were included in the present research. The light-time effect was adopted as a main factor for the
detailed description of the long-term period changes. Third bodies were found with orbital periods from 46 up to
84~years, and eccentricities from 0.0 to 0.78 for the selected binaries. The mass functions and the minimal
masses of such bodies were also calculated.


\section{Introduction}

Eclipsing Binaries (hereafter EBs) are excellent objects for
determining the physical properties of stars and detecting
additional components in these systems. The long-time behavior of
the period of an EB could reveal the presence of another component
orbiting with the EB around the common center of mass. Photometric
observations of EBs sometimes cover more than a century, so the
periods of such third bodies could reach the same time interval.

The motion around the barycenter causes apparent changes of the
observed binary period with a period corresponding to the orbital
period of the third body, called the LIght-Time Effect (or
'light-travel time', hereafter LITE). \cite{Irwin1959} improved
the method developed by \cite{Woltjer1922} for analyzing the
long-term variation of the times of minima caused by the third
body orbiting the eclipsing pair. Useful comments and limitations
were discussed by \cite{FCH73} and by \cite{Mayer1990}. Nowadays
there are more than one hundred EBs showing LITE, where the effect
is certainly presented or supposed (see e.g.
\cite{BorkovitsHegedus}, \cite{Albayrak1999}, \cite{Wolf2004},
\cite{Hoffman2006}, etc.). For the apparent period changes in EBs
according to their $O-C$ diagrams, see e.g. the catalogue by
\cite{Kreiner2001}. From the same paper the look of the $O-C$
diagrams, presented in the present paper, was adopted. In the
figures \ref{FigDKCep} to \ref{Figs} of the present work the full
circles represent the primaries and the open circles the
secondaries, the bigger the point, the bigger the weight. For the
limitations and consequences of the $O-C$ diagram analysis, see
e.g. \cite{Sterken2005}.

The computation of the parameters of the third-body orbit is a
classical inverse problem with 5 parameters to be found -- $p_3$,
$T_0$, $A$, $\omega$, $e_3$, which indicate the period of the
third body, the periastron passage, the semi-amplitude of the
light-time effect, the argument of periastron and the
eccentricity, respectively (for a detailed description see e.g.
\citealt{Mayer1990}). The ephemerides for the individual system
($JD_0$ and $P$ for the linear one and $q$ for the quadratic one)
have to be calculated together with the parameters of LITE. The
mass function $f(M_3)$ and the minimal mass of the third component
$M_{3,min} = M_3 \cdot \sin i_3$ (for $i_3 = 90^\circ$) could be
computed from this set of parameters. The weights assigned to
individual observations were used as following: $w=1$ for visual
observations, 3-5 for photographic and 10 for CCD and
photoelectric observations. The computing code itself could be
downloaded via the web pages of the
author\footnote{\href{http://sirrah.troja.mff.cuni.cz/~zasche/}{http://sirrah.troja.mff.cuni.cz/$\sim$zasche/}}.

All of the selected systems are Algol-type EBs, semi-detached and
also their spectral types are similar. The primary components have
spectral types from B to F, while the secondaries from G to K.
Except for 14 observations in Table \ref{Minima}, all the times of
minima used in this paper were collected from the published
literature and from minima databases available in the internet.

According to a recent paper on the period changes in Algols by
\cite{Hoffman2006}, there could be a connection between the
spectral type of the secondary component and the nature of the
period changes. Systems with spectral types of secondaries later
than F5 show $O-C$ variations, which could be caused by the
magnetic activity cycles and convective envelopes. This effect was
discussed by \cite{Hall1989}, \cite{Applegate1992},
\cite{Lanza1998}, etc. The role of magnetic cycles on the period
changes is discussed below, but due to lack of information about
the systems such an analysis is a difficult task. For some of the
systems selected in this paper the spectral types of secondaries
are known with a low confidence level, light-curve analysis is
missing and spectroscopy has never been done.

\section{Observations}

The photometric observations were obtained by CCD detectors at two
different observatories.

The first one is situated in Ond\v{r}ejov Observatory, Czech
Republic. The 65-cm telescope was equipped with the Apogee AP-7
CCD camera. All of the observations were secured during 2006 and
2007 and only $R$ filter was used. The exposure times depend on
the particular target and observing conditions, ranging from 20 to
60 seconds.

\begin{table}[b!]
\caption{New times of minima based on CCD observations (Kwee-van
Woerden (\citeyear{Kwee}) method was used). All of them were
obtained in $R$ filer. For the explanation of the observatory
abbreviations see the text.} \label{Minima} \centering
\begin{tabular}{c c c c c }
\hline
 Star   & HJD-2400000 & Error  & Type & Observatory \\
\hline
 DK~Cep & 54348.35978 & 0.00010 &  II & Ath \\
 DK~Cep & 54372.51532 & 0.00008 &  I  & Ath \\
 DK~Cep & 54376.45890 & 0.00005 &  I  & Ath \\
 TY~Del & 54252.56885 & 0.00020 &  I  & Ond \\
 TZ~Eri & 54368.55806 & 0.00008 &  I  & Ath \\
 RR~Dra & 54203.36519 & 0.00003 &  I  & Ond \\
 VX~Lac & 54350.36051 & 0.00006 &  II & Ath \\
 VX~Lac & 54359.49220 & 0.00003 &  I  & Ath \\
 VX~Lac & 54373.46069 & 0.00003 &  I  & Ath \\
 UZ~Sge & 54035.31348 & 0.00028 &  I  & Ond \\
 UZ~Sge & 54252.45613 & 0.00019 &  I  & Ond \\
 UZ~Sge & 54293.45120 & 0.00023 &  II & Ath \\
 UZ~Sge & 54313.39219 & 0.00021 &  II & Ath \\
 UZ~Sge & 54314.49758 & 0.00013 &  I  & Ath \\
\hline
\end{tabular}
\end{table}

The second observatory is situated at Athens University campus,
Athens, Greece. The 40-cm telescope is equipped with the SBIG
ST-8XMEI CCD camera. All of the observations were obtained during
2007, using only the $R$ filter. The exposure times depend on the
individual stars and observing conditions, ranging from 15 to 120
seconds.

\section{Analysis of individual systems}

\subsection{DK~Cep}

The first system is the eclipsing binary DK~Cep. This neglected
system is about 12.2 mag bright in $B$ filter. The spectrum was
classified as A8 + G4IV in \cite{Svechnikov1990}, but only on the
basis of photometric indices. Its photometric variability was
discovered in 1966 by \citeauthor{ObjevDKCep1966}, but since then
no analysis of this system was performed (neither photometric nor
spectroscopic one). Also period analysis has not been performed so
far.

Altogether 157 observations of minimum light were collected for
the analysis. One secondary and two primary times of minima were
also observed at Athens observatory (see Table \ref{Minima}).

\begin{figure}[t!]
\hspace{1.5cm} \includegraphics[width=14cm]{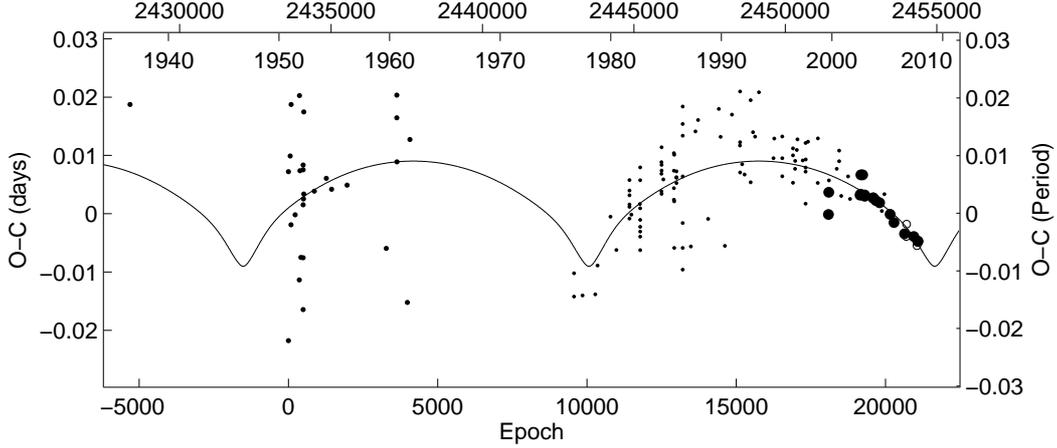}
 \caption{The $O\!-\!C$ diagram of DK~Cep. The individual observations
 are shown as dots (primary) and open circles (secondary), the small
 ones for visual and the large ones for CCD and photoelectric observations,
 the bigger the symbol, the bigger the weight. The curve represents the
 predicted LITE variation (see the text).
 \label{FigDKCep}}
\end{figure}

The analysis was carried out and the resultant plot is given in
Fig.\ref{FigDKCep}, where the theoretical curve is shown together
with the individual observations. The parameters computed from our
analysis are given in Table \ref{TableBig1}. Regrettably, the
phase near the periastron is not covered with the data. This
affects the values of $\omega$ and $e$ and their respective
errors, so the orbit could have slightly different eccentricity.
This hypothesis could be proved in the upcoming periastron
passage, which will occur near 2010.

From the parameters of the third-body orbit, the predicted minimal
mass of this component can be also derived. Using the masses of
the primary and secondary of the EB pair as $M_1 = 1.65$\Mo~and
$M_2 = 0.92$\Mo~(according to \citealt{Svechnikov1990}), the
minimal mass results in $M_{3,min} = 0.32$\Mo. Due to large
difference between this value and the values of masses of the
eclipsing pair components, one could speculate about the possible
detectability of such a body. This mass indicates a spectral type
about M3 (according to \citealt{Hec1988}). The magnitude
difference between the primary and the tertiary component is
therefore more than 6~mag, or the amount of the third light in the
light curve less than 1 percent, which is undetectable. Also, in
the spectrum of the system such a body will be, probably, hardly
detectable. Since the distance to the system is not known, one
cannot estimate the predicted angular separation of the third
component.

\begin{table*}[t]
\small \caption{The final results (part 1): Parameters of the
third-body orbits from the analysis of times of minima. The table
is divided into two parts. In the first one the computed
parameters are presented, while in the second one the derived
quantities are given (the mass $M_{12} = M_1 + M_2$ is taken from
the literature).} \label{TableBig1} \centering \scalebox{0.85}{
\begin{tabular}{c c c c c c c }
\hline
 Parameter         &           DK~Cep         &       TY~Del             &       RR~Dra           \\
 \hline
 $JD_0$ $[$HJD$]$  & $2433590.556 \pm 0.005$  & $2442959.471 \pm 0.001$  &  $2434913.728 \pm 0.022$  \\
 $P$ $[$day$]$     & $0.9859085 \pm 0.0000004$&$1.1911264 \pm 0.0000002$ & $2.8312140 \pm 0.0000053$ \\
 $p_3$ $[$yr$]$    &    $31.3 \pm 2.5$        &    $64.9 \pm 2.3$        &  $84.3 \pm 0.6$            \\
 $T_0$ $[$HJD$]$   &    $2454900 \pm 1200$    &  $2449192 \pm 1017$      &  $2450058 \pm 442$         \\
 $\omega$ $[$deg$]$&     $273 \pm 32$         &   $38.5 \pm 18.2$        &  $110.2 \pm 3.5$           \\
 $e$               &   $0.780 \pm 0.256$      &   $0.221 \pm 0.055$      &  $0.503 \pm 0.028$         \\
 $A$ $[$day$]$     &   $0.009 \pm 0.002$      &   $0.0268 \pm 0.0012$    &  $0.0731 \pm 0.0017$        \\
 $q$ $[$day$]$     &          --              &      --                  &$(-126.2 \pm 0.1) \cdot 10^{-10}$ \\
 \hline
 $M_{12}$ [\Mo]    &        $2.57$            &     $3.64$               &  $2.75$                  \\
 $f(M_3)$ [\Mo]    &  $0.0039 \pm 0.0002$     &  $0.025 \pm 0.001$       &  $0.300 \pm 0.002$    \\
 $M_{3,min}$ [\Mo] &   $0.32 \pm 0.01$        &    $0.79 \pm 0.07$       &  $1.85 \pm 0.09$     \\
 \hline
\end{tabular}}
\end{table*}

\subsection{TY~Del}

TY~Del (AN~141.1935, BD+12~4539) is another EB with apparently
variable period. The spectrum was classified as B9+G0IV
(\citealt{Hoffman2006}) and the relative brightness about 10.1 mag
in $V$ filter. There is a consensus about the spectral types of
the components of TY~Del, but there is a difference between the
masses. \cite{Brancewicz1980}, and also \cite{Budding1984} and
\cite{Budding2004} have proposed masses $M_1 = 5$\Mo, $M_2 =
2$\Mo, while \cite{Svechnikov1990} gave $M_1 = 2.8$\Mo, $M_2 =
0.84$\Mo. The latter are more likely to the proposed spectral
types.

The star was discovered to be a variable by
\cite{Hoffmeister1935}. There was only one attempt to observe the
whole light curve of TY~Del photoelectrically by
\cite{Faulkner1983}, unfortunately only about a half of the curve
was observed. No analysis of these data was carried out so far.
The star was also studied by \cite{Cook1993} on the basis of his
visual observations for the long-time scale intrinsic variations,
but the results were not very conclusive.

The spectroscopic observations in H$\alpha$ were obtained by
\cite{Vesper2001}. They concluded that there was no activity in
H$\alpha$ and no evidence for the mass transfer structures was
found in this system.

\begin{figure}
 \hspace{1.5cm}
 \includegraphics[width=14cm]{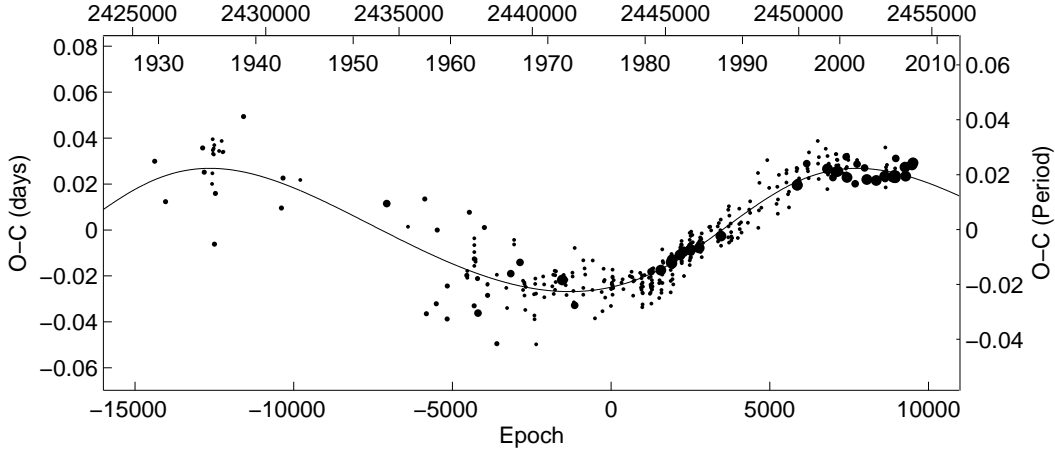}
 \caption{The $O-C$ diagram of TY~Del (for the description see Fig.\ref{FigDKCep}).}
 \label{FigTYDel}
\end{figure}

Altogether 368 times of minima were collected, from which only 5
were omitted due to their large scatter. One period of the third
body is already sufficiently covered by the data points (see Fig.
\ref{FigTYDel}), but further observations are still needed. From
the LITE parameters (see Table \ref{TableBig1}) and with the
approximate masses of the EB components as $M_{12} = 3.64$\Mo~
\citep{Svechnikov1990} it was possible to derive the minimal mass
of the third component as $M_{3,min} = 0.67$\Mo. Due to lack of
any other observations, this hypothesis cannot be proved. The
spectral types and masses were derived only on the basis of the
photometric indices and are not very conclusive. A spectroscopic
analysis, as well as an analysis of the light curve of the system
is needed, but the third light is hardly detectable in the
light-curve solution. Regrettably, the star was not measured by
Hipparcos, so the distance is not known and one cannot derive the
predicted angular separation of the third component. There is also
clearly visible some additional short-periodic variation, which
could not be described by the LITE. For a discussion about its
possible explanation see Section \ref{AlternativeExpl}.

\subsection{RR~Dra}

Another eclipsing binary which exhibits apparent period changes is
RR~Dra (AN~188.1904). Its relative brightness is about 9.8 mag in
$V$ filter, and the spectrum was classified as A2+G8IV
\citep{Svechnikov1990}, while \cite{Yoon1994} proposed a later
spectral type K0 of secondary. \cite{Svechnikov1990} gave the
masses $M_1 = 2.15$\Mo~ and $M_2 = 0.6$\Mo. The primary minimum is
very deep, about 3.5 mag and the orbital period is about 2.8 days.

\begin{figure}[t!]
 \hspace{1.5cm}
 \includegraphics[width=14cm]{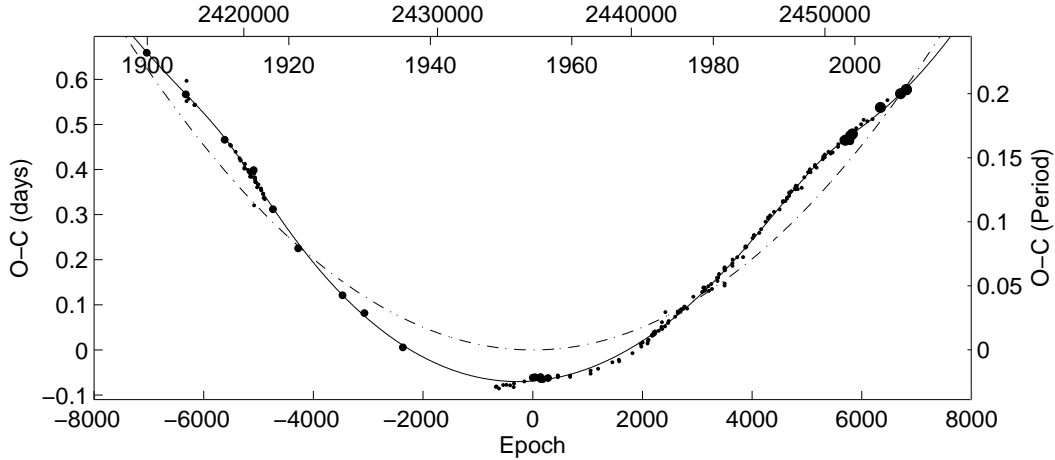}
 \caption{The $O-C$ diagram of RR~Dra (for the description see Fig.\ref{FigDKCep}).}
 \label{FigRRDra1}
\end{figure}

\begin{figure}[b!]
 \hspace{1.5cm}
 \includegraphics[width=14cm]{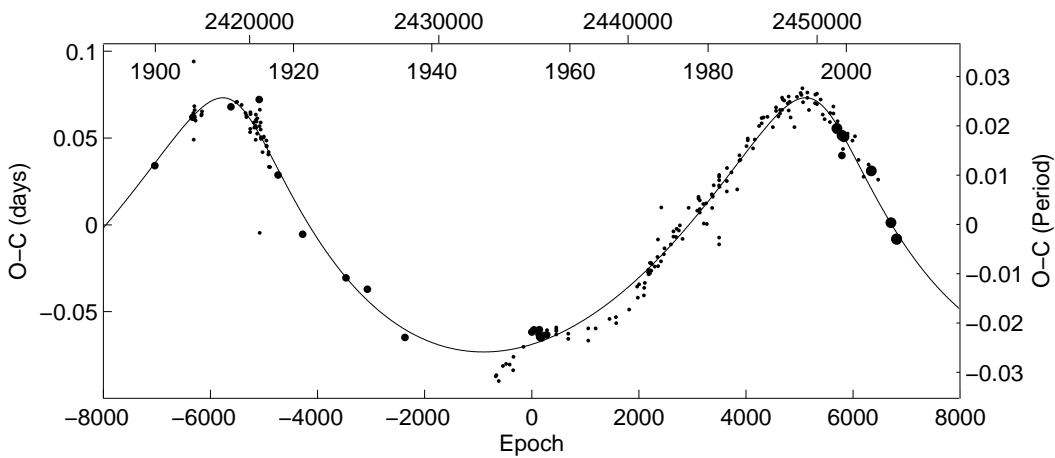}
 \caption{The $O-C$ diagram of RR~Dra after subtraction of the quadratic
 term.
 \label{FigRRDra2}}
\end{figure}

The star was discovered to be a variable by \cite{Ceraski1905}. The minimum is so deep that visual observers
could also provide reliable observations. That is the reason why most of the collected times of minima are
visual ones (193 out of 219). \cite{Kreiner1971} collected all available minima for the period analysis. The
long-term increase of the period, evident from his $O-C$ diagram, is very probably due to the mass transfer
between the components. The most recent period study of this system was performed by \cite{Qian2002}, who
considered (besides the mass transfer) the abrupt period jumps - altogether 8 jumps were introduced to describe
the $O-C$ diagram in detail. Almost the same goodness of fit could be reached by applying the LITE hypothesis
besides the mass transfer.

A total of 219 times of minima were used for the present analysis.
The $O-C$ plot is presented in Fig.\ref{FigRRDra1}, where LITE and
the quadratic term were plotted together. In Fig.\ref{FigRRDra2}
only LITE is shown. It is obvious that the period increase is very
rapid, and the amplitude of LITE is still quite high. This leads
to the relatively high mass function, which results in high
predicted minimal mass $M_{3,min} = 1.85$\Mo. This is larger than
the secondary and should be probably observable in the light-curve
solution as well as in the spectrum. Regrettably, no such analysis
has been performed so far.

The quadratic term coefficient $q = (126.1 \pm 0.1) \cdot
10^{-10}$~day leads to a period change about $3.25 \cdot
10^{-6}$~day/yr. From this value the conservative mass transfer
rate could be derived $\dot M = 3.2 \cdot 10^{-7}$~\Mo/yr. This
relatively high value of mass transfer rate arises from the very
rapid period change, which was attributed to the quadratic
ephemeris. The spectroscopic observations during the primary
eclipse made by \cite{Kaitchuck1985} indicate a possible presence
of a transient accretion disc in the system. The presence of such
a disc also supports the hypothesis of mass transfer in the
system.

For the estimation of the angular separation of the third body and
the astrometric confirmation of the LITE hypothesis, the distance
to the system has to be known. The star was not measured by
Hipparcos and the distance is not known precisely. The only
information about the distance comes from \cite{Kharchenko2001},
where is surprisingly introduced an inaccurate value of the
parallax $\pi = (0.40 \pm 11.50)$~mas. A distance with such a
large error is useless for the estimation of the angular
separation of the predicted component.

\subsection{TZ~Eri}

The system TZ~Eri (AN~40.1929, BD-06~880) is an EB with orbital
period of about 2.6 days and apparent brightness of about 9.7~mag
in $V$ filter. It has a deep primary minimum (about 2.8 mag), so
the visual observations could be also reliable.

\begin{figure}[t!]
 \hspace{1.5cm}
 \includegraphics[width=14cm]{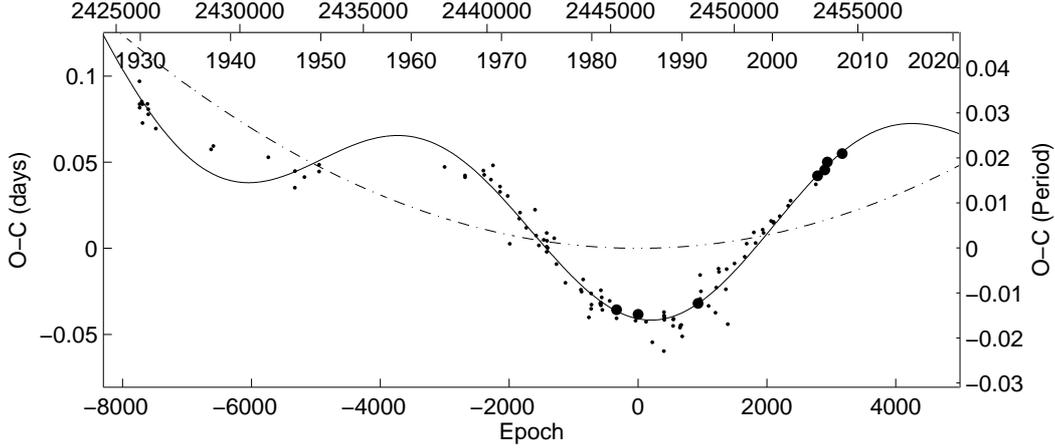}
 \caption{The $O-C$ diagram of TZ~Eri (for the description see Fig.\ref{FigDKCep}).}
 \label{FigTZEri}
\end{figure}

Its variability was discovered by \cite{Hoffmeister1929}, who recognized the system to be of Algol-type. The
spectral type was firstly classified by \cite{Cannon1934} as F. Later, the spectrum was re-classified by
\cite{Barblan1998} as A5/6 V (primary) and K0/1~III (secondary). In this latter paper the light-curve
observations in the Geneva 7-color photometric system were analyzed together with the radial-velocity curves of
both components. The Wilson-Devinney code was used, resulting in a set of parameters describing both components.
For our analysis the masses $M_1 = 1.97$\Mo, $M_2 = 0.37$\Mo~were used \citep{Barblan1998}.

Several studies about the presence of an accretion disc in the system have also been made by
\cite{Kaitchuck1982}, \cite{Kaitchuk1988}, \cite{Vesper2001} and others. This disc as well as mass transfer from
the secondary to the primary is in agreement with our result about the increasing orbital period (see below).
The system was also included in the sample of Algol-type binaries with radio emission \citep{Umana1998}.
Recently, the star was investigated according to a possible connection between the orbital and pulsational
periods (see \citealt{Soydugan2006}).

The analysis of the long-term period changes was based on a set of
108 observations (mostly the visual ones). The resultant $O-C$
diagram is presented in Fig.\ref{FigTZEri} and the parameters of
the predicted LITE are given in Table \ref{TableBig2}. The minimal
mass of the third component results in $M_{3,min} = 1.3$\Mo, or
the spectral type F6 (according to \citealt{Hec1988}). Such a body
could be detected in the light-curve solution as well as in the
spectra of TZ~Eri. Regrettably, there was no attempt to detect
such a body during the detailed analysis by \cite{Barblan1998}.
The long-term period increase is due to the mass transfer from the
secondary to the primary, with the conservative mass-transfer rate
$\dot M = 3.2 \cdot 10^{-8}$\Mo/yr.

\begin{table*}[t!]
 \small \caption{The final results, part 2. The distance $d$ of the system TZ~Eri is taken from the literature. } \label{TableBig2}
 \centering
 \scalebox{0.85}{
 \begin{tabular}{c c c c c c c }
 \hline
 Parameter         &        TZ~Eri            &          VX~Lac          &        UZ~Sge             \\
 \hline
 $JD_0$ $[$HJD$]$  & $2446109.730 \pm 0.009$  &  $2440908.901 \pm 0.001$ & $2445861.420 \pm 0.002$   \\
 $P$ $[$day$]$     & $2.6061129 \pm 0.0000034$& $1.0744976 \pm 0.0000002$& $2.2157425 \pm 0.0000007$ \\
 $p_3$ $[$yr$]$    & $48.8 \pm 6.8 $          &      $68.2 \pm 0.3$      & $47.0 \pm 2.4$            \\
 $T_0$ $[$HJD$]$   & $2451107 \pm 2290$       &    $2440000 \pm 1100$    & $2449248 \pm 2036$       \\
 $\omega$ $[$deg$]$&  $0.0 \pm 40.4$          &       $111 \pm 15$       & $293.6 \pm 46.7$        \\
 $e$               &  $0.005 \pm 0.105 $      &     $0.41 \pm 0.12$      & $0.281 \pm 0.130$      \\
 $A$ $[$day$]$     &  $0.0423 \pm 0.0138 $    &   $0.0210 \pm 0.0016$    & $0.0235 \pm 0.0029$   \\
 $q$ $[$day$]$     & $(18.0 \pm 0.2)\cdot 10^{-10}$& $(3.15 \pm 0.01)\cdot 10^{-10}$ &  --     \\
 \hline
 $M_{12}$ [\Mo]    &  $2.34$                  &         $1.92$           & $2.34$          \\
 $d$ [pc]          &  $270 \pm 12$            &           --             &   --          \\
 \hline
 $f(M_3)$ [\Mo]    &  $0.165 \pm 0.013$       &   $0.0108 \pm 0.0003$    & $0.031 \pm 0.002$  \\
 $M_{3,min}$ [\Mo] &  $1.30 \pm 0.10$         &    $0.385 \pm 0.008$     & $0.65 \pm 0.05$   \\
 $a$ [mas]         &  $77 \pm 7$              &          --              &  --         \\
 \hline
 \end{tabular}}
\end{table*}

Despite the fact that the star was not observed by Hipparcos,
\cite{Barblan1998} estimated the photometric distance to $d = (270
\pm 12)$~pc. Assuming a coplanar orbit of the third component,
then $M_3 = M_{3,min}$ and one could calculate the predicted
angular separation of the third body to $a = 77$~mas and a
magnitude difference between the eclipsing pair and the predicted
component about 1.7~mag. Therefore, the final result is that such
a body is detectable with the modern stellar interferometers.

\subsection{VX~Lac}

The eclipsing binary system VX~Lac (BD +37 4662, GSC 03214-01295)
has its apparent brightness about 10.55~mag in $V$ filter. The
orbital period is close to one day. Its spectral type was firstly
derived by \cite{Cannon1934} as F0, nowadays the adopted spectral
type is F0 + K4IV \citep{Svechnikov1990}.

\begin{figure}[t!]
 \hspace{1.5cm}
 \includegraphics[width=14cm]{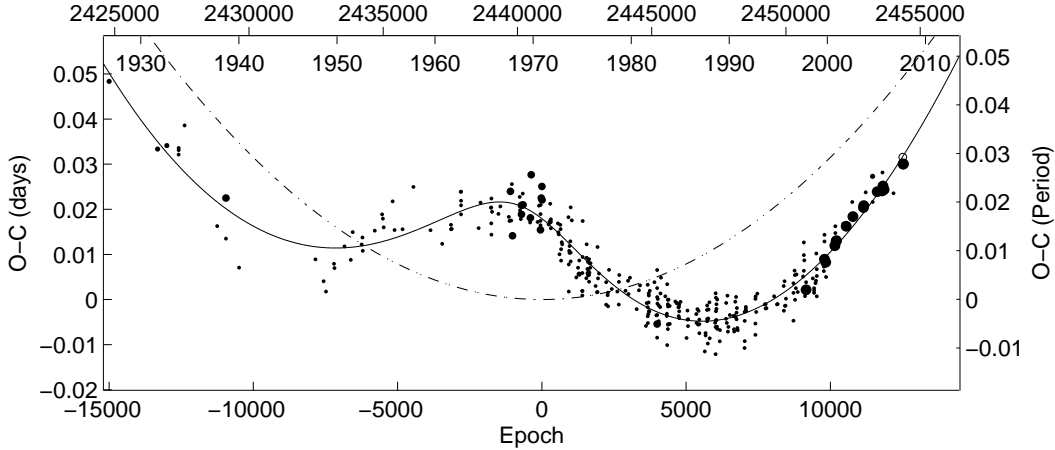}
 \caption{The $O-C$ diagram of VX~Lac (for the description see Fig.\ref{FigDKCep}).}
 \label{FigVXLac}
\end{figure}

Unfortunately, neither light curve analysis nor spectroscopic mass
ratio exists for the system. \cite{Kreiner1971} included this star
to his study of systems with period changes, but at that time the
LITE variation could not be identified conclusively. Nowadays
there are 370 times of minima. It is clearly visible from this
data set that, besides the quadratic term, there is a periodic
variation in the times of minima (see Fig.\ref{FigVXLac}). The
parameters of the fit are given in Table \ref{TableBig2}.

The mass transfer coefficient leads to the conservative mass
transfer rate about $\dot M = 4.7 \cdot 10^{-8}$\Mo/yr. Such a
result is in agreement with the previous study by \cite{Chis1998},
who concluded that the star is currently in the mass loss state.

The LITE parameters indicates the predicted minimal third-body
mass about $0.4$\Mo. The secondary component ($M_2 = 0.47$\Mo) is
only slightly more massive, but, despite this fact, the expected
magnitude difference between the primary and the tertiary
component is too high (about 5~mag) to be detected. The derived
contribution of the third star to the total luminosity is only
about 1 percent and therefore is hardly detectable. Having no
information about its distance one cannot estimate the angular
separation of the third body. The additional variation in $O-C$
diagram after subtracting the LITE and quadratic term is shown in
Fig.\ref{Figs} (see the possible explanation in Section
\ref{AlternativeExpl}).

\subsection{UZ~Sge}

The EB system UZ~Sge (AN~435.1936) has orbital period of about 2.2
days and its spectrum is classified as A3V+G0IV
\citep{Svechnikov1990}.

The photometric variability of UZ~Sge was discovered by \cite{Guthnick1939}. Since then, there was no attempt
for any detailed analysis, neither photometric nor spectroscopic one. The only spectroscopic observation was
done by \cite{Halbedel1984} for the derivation of the spectral type of primary component.

\begin{figure}[t!]
 \hspace{1.5cm}
 \includegraphics[width=14cm]{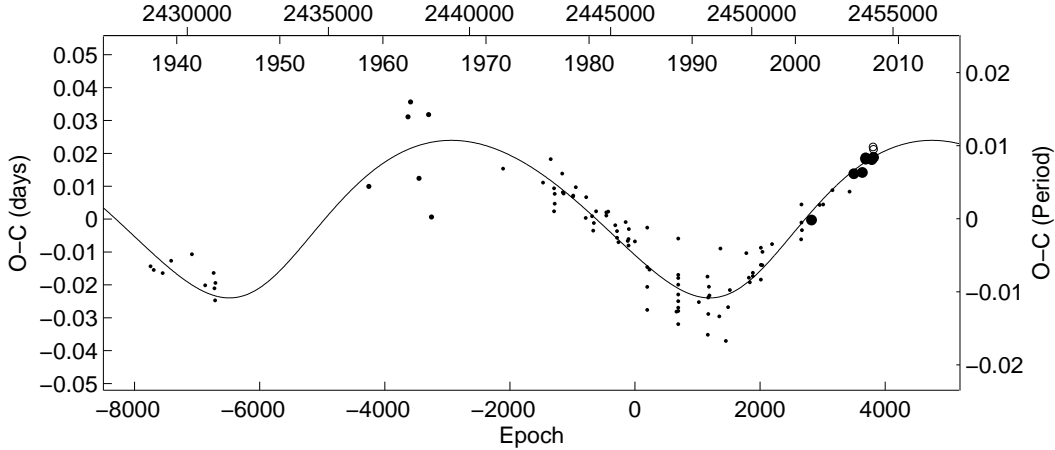}
 \caption{The $O-C$ diagram of UZ~Sge (for the description see Fig.\ref{FigDKCep}).}
 \label{FigUZSge}
\end{figure}

Five new times of minima were obtained for this paper (see Table~\ref{Minima}). Altogether 125 measurements of
times of minima were found in literature, but 14 observations were neglected due to their large scatter. Taking
as masses of the individual components those given by \cite{Svechnikov1990}, $M_1 = 2.05$\Mo~and $M_2 =
0.29$\Mo, the minimal mass of the third component results in $M_{3,min} = 0.65$\Mo. But due to absence of any
detailed analysis of this system, this value cannot be proved.

\subsection{Alternative explanation of the orbital period changes} \label{AlternativeExpl}

\begin{figure}[t!]
 \hspace{1.5cm}
 \scalebox{0.77}{\includegraphics[15mm,190mm][30mm,270mm]{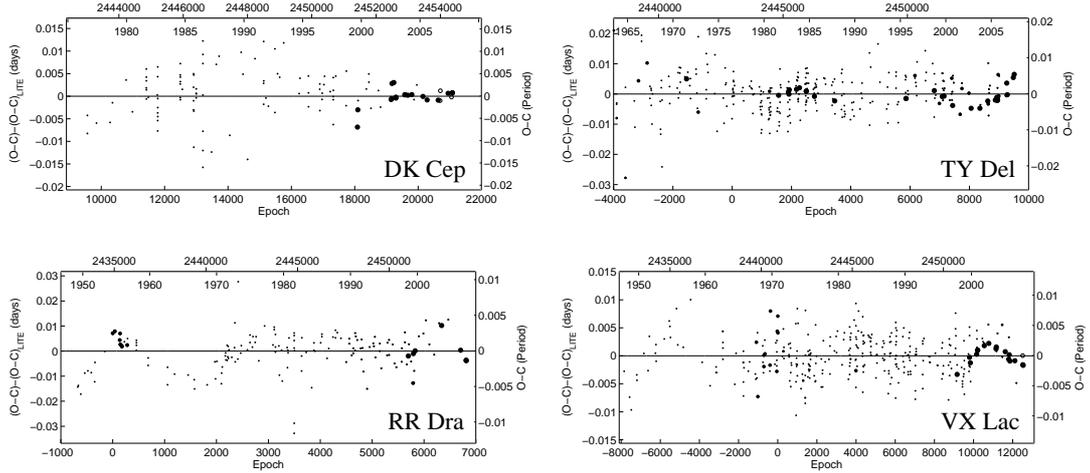}}
 \caption{The $O-C$ diagrams after subtraction of LITE.}
 \label{Figs}
\end{figure}

One can see some additional non-periodic variations in the $O-C$
diagrams, which cannot be described by applying only the LITE
hypothesis. In Fig. \ref{Figs} four cases with the most evident
variations are shown. The amplitudes of these variations are
usually about 10 minutes in $O-C$ diagram and these are not
strictly periodic. This could be caused by the presence of stellar
convection zones, in an agreement with the so-called Applegate's
mechanism, see e.g. \cite{Applegate1992}, \cite{Lanza1998}, or
\cite{Hoffman2006}. This effect could play a role, because the
spectral types of the secondary components are later than F5 (see
\cite{Zavala2002} for a detailed analysis). To conclude, for a
better description of the observed period variations of these
systems, the magnetic activity cycles could be present together
with the LITE. On the other hand, one has to take into
consideration that the spectral types of most of these binaries
were derived only on the basis of their photometric indices
\citep{Svechnikov1990} and are not very reliable.

The only system, where one could estimate the variation of the
quadruple moment (see \citealt{Applegate1992}) required to explain
the long-term period variations, is TZ~Eri, where the semi-major
axis of the orbit from the light curve and radial velocity curve
solution is known. Using the following equation $$ \Delta P = A
\sqrt{2 (1 - \cos \{ 2 \pi P/p_3 \} )} $$ \citep{Rovit2000} one
could compute the amplitude of the period oscillation. The period
variation $\Delta P / P$ can be used for calculating the variation
of the quadruple moment $\Delta Q$, using the equation
\citep{Lanza2002} $$ \frac{\Delta P}{P} = -9 \frac{\Delta
Q}{Ma^2}.$$ This quantity results in $\Delta Q = 3.6 \cdot
10^{51}$~g$\cdot$cm$^2$, which is well within the limits for
active binaries and therefore the variation in TZ~Eri could be
also explained by this mechanism.

\section{Discussion and conclusions}

Six Algol-type semi-detached eclipsing binaries were analyzed for
the presence of LITE on the basis of their $O-C$ diagram analysis
and the times-of-minima variations. A few new observations of
these systems were obtained and used in the present analysis. All
of the studied systems show apparent changes of their orbital
periods, which could be explained as a result of orbiting the EB
with a third component around their common center of mass.

Such a variation has usually a period of the order of decades, as
one can see from Figs.\ref{FigDKCep}--\ref{FigUZSge}, which can be
described by applying the LITE hypothesis sufficiently. In three
cases (RR~Dra, TZ~Eri and VX~Lac) the quadratic term in the light
elements was also used. This could be explained as a mass transfer
between the two components, which is a common procedure in
semi-detached systems. The conservative mass-transfer rates were
calculated.

About half of the systems show also an additional variation in
$O-C$ diagram after the subtraction of LITE. Such a variation is
not strictly periodic, it has usually an amplitude about 10
minutes and a "period" from 5 to 20 years. Because all of the
systems have the spectral types of secondaries later than F5, the
variation could be caused by the convection and the magnetic
activity in these binaries. This explanation would clarify the
non-periodicity and the changes in amplitude of such variations,
as well the existence or non existence of such phenomena in some
EBs.

Regrettably, in most of the systems no detailed analysis (neither
photometric nor spectroscopic) has been made so far. The spectral
types and the masses of the individual components in the systems
are only approximate, so the parameters of the predicted third
bodies are also affected by relatively large errors. Due to
missing information about the distances to these binaries, any
prediction about the angular separation could be done. It is
obvious that only further detailed photometric, as well as
spectroscopic and interferometric analysis would reveal the nature
of these systems and confirm or reject the third-body hypothesis.

\medskip
\medskip
\medskip
\medskip

\noindent {\bf \underline{Acknowledgments:}} This investigation was supported for P.Z. \& M.W. by the
Czech-Greek project of collaboration No. 7-2006-5 {\it Photometry and Spectroscopy of Binaries} of Ministry of
Education, Youth and Sport of the Czech Republic, and for A.L. \& P.N. by the Special Account for Research
Grants 70/3/8680 of the National \& Kapodistrian University of Athens, Greece. This research has made use of the
SIMBAD database, operated at CDS, Strasbourg, France, and of NASA's Astrophysics Data System Bibliographic
Services.

\end{document}